\documentclass{iopart}
\usepackage[utf8]{inputenc}
\usepackage{cmap}
\usepackage[T1]{fontenc}

\usepackage{graphicx}

\expandafter\let\csname equation*\endcsname\relax
\expandafter\let\csname endequation*\endcsname\relax
\usepackage{amsmath}
\usepackage{amsthm, amssymb, color}
\usepackage{sistyle}
\usepackage{multirow}

\newcommand{\eps}{\varepsilon}

\newcommand{\lam}{\lambda}

\newcommand{\D}{{\rm d}}

\newcommand\imeq{\overset!=}

\makeatletter
 \newcommand\kla[1]{\left(#1\right)}               
 \newcommand\skal[1]{\left\langle#1\right\rangle}  
\makeatother

\usepackage[acronym]{glossaries}
\newacronym{kpz}{KPZ}{Kardar--Parisi--Zhang}
\newacronym{ew}{EW}{Edwards--Wilkinson}
\newacronym{rsos}{RSOS}{restricted solid-on-solid}
\newacronym{bd}{BD}{ballistic deposition model}
\newacronym{ms}{MS}{multisurface coding}
\newacronym{fdr}{FDR}{fluctuation-dissipation relation}
\newacronym{kk}{KK}{Kim--Kosterlitz}
\newacronym{mcs}{MCS}{Monte-Carlo steps}

\newacronym{rng}{RNG}{random number generator}
\newacronym{gpu}{GPU}{graphics processing unit}
\newacronym{gpgpu}{GPGPU}{general purpose computing on graphics processing units}
\newacronym{cpu}{CPU}{central processing unit}
\newacronym{simt}{SIMT}{single instruction multiple thread}

\newacronym{dd}{DD}{domain decomposition}
\newacronym{cb}{CB}{checker-board}
\newacronym{sca}{SCA}{stochastic cellular automaton}
\newacronym{rs}{RS}{random-sequential}
\newacronym{dt}{DT}{double tiling}
\newacronym{db}{DB}{dead border}
\newacronym{cdb}{cDB}{coarse dead border}
\newacronym{dtr}{DTr}{\gls{dt} \gls{dd} with random origin}
\newacronym{dtrdt}{DTrDT}{\gls{dtr} at device level and single-hit
\gls{dt} at block level}
\newacronym{dtrdb}{DTrDB}{\gls{dtr} at device level and single-hit
\gls{db} at block level}
\newacronym{dtrdtr}{DTrDTr}{\gls{dtr} at device level and single-hit
\gls{dtr} at block level}

\newacronym{tc}{TC}{thread cell}
\newacronym{t}{T}{thread}

\newacronym{rhs}{r.h.s.}{right-hand side}
\newacronym{resp}{resp.}{respectively}
\newacronym{snr}{S/N}{signal-to-noise ratio}
\newacronym{fom}{FOM}{figure of merit}

\newacronym{lsi}{LSI}{local scale-invariance}

\glsunset{gpu}
\glsunset{cpu}

\makeglossaries

\newcommand\NkpzArScaSXXX{23849}
\newcommand\NkpzArScaSC{12012}

\begin{document}

\title{Local Scale-Invariance of the 2+1 dimensional Kardar--Parisi--Zhang model}

\author{Jeffrey Kelling\textsuperscript{2,3}, G\'eza \'Odor\textsuperscript{1}, and Sibylle Gemming\textsuperscript{3,4}}

\address{\textsuperscript1Institute of Technical Physics and Materials Science,
Centre for Energy Research of the Hungarian Academy of Sciences \\
P.O.Box 49, H-1525 Budapest, Hungary \\}
\address{\textsuperscript2Department of Information Services and Computing, \\
Helmholtz-Zentrum Dresden-Rossendorf \\
P.O.Box 51 01 19, 01314 Dresden, Germany\\}
\address{\textsuperscript3Institute of Ion Beam Physics and Materials Research \\
Helmholtz-Zentrum Dresden-Rossendorf \\
P.O.Box 51 01 19, 01314 Dresden, Germany\\}
\address{\textsuperscript4Institute of Physics, TU Chemnitz\\
09107 Chemnitz, Germany}

\begin{abstract}
Local Scale-Invariance theory is tested by extensive
dynamical simulations of the driven dimer lattice gas model,
describing the surface growth of the 2+1 dimensional
Kardar--Parisi--Zhang surfaces. Very precise measurements of
the universal autoresponse function enabled us to perform
nonlinear fitting with the scaling forms, suggested by \gls{lsi}.
While the simple \gls{lsi} ansatz does not seem to work, forms based
on logarithmic extension of \gls{lsi} provide satisfactory description of
the full (measured) time evolution of the autoresponse function.
\end{abstract}
\maketitle

Understanding universal scaling behavior of nonequilibrium dynamical
systems is a challenging task \cite{odorbook}.
Critical phenomena can emerge away from equilibrium, but due to the
broken time reversal and translational symmetries, an extension of
the Renormalization Group method (RG), as the best tool, is not straightforward \cite{Tauber_2014_book}.
The lack of translational symmetry manifests in aging phenomena observed in
glasses, polymers, reaction-diffusion systems or cross-linked
networks \cite{St78}.

\gls{lsi} theory is proposed \cite{HLSI94} to generalize dynamical
scaling to a larger set of local scale
transformations, including $t\to t/(1+ t \gamma)$, analogously
as conformal invariance (CI) extends RG of equilibrium critical
phenomena.
As CI~\cite{CI,HCI} works well in case of equilibrium universality classes,
\gls{lsi} aims at the same for nonequilibrium dynamical ones~\cite{HP}.
\gls{lsi} has been shown to reproduce the universal shapes of responses
and correlators in a large variety of models, as reviewed
in detail in \cite{HP}. The predictive power of generalized dynamical scaling
alone was shown to be limited~\cite{HinLSI}, and later the role of the generalized
Galilei invariance was recognized.
Analogously to the logarithmic CI generalization \cite{LCI},
Henkel suggested the logarithmic extension of \gls{lsi} (LLSI) to make the
theory applicable for more general cases \cite{Henkel2013}.

While many systems are described by a single dynamical length scale
$L(t)\sim t^{1/z}$, with the dynamical exponent
$z$~\cite{Hohen,Bray94},
aging ones are best characterized by two-time quantities,
such as the dynamical correlation and response functions \cite{Cug}.
In the aging regime: $s\gg \tau_{\rm m}$ and $t-s\gg \tau_{\rm m}$,
where $\tau_{\rm m}$ is a microscopic time scale, one expects the
following law for the autoresponse function of the field $\phi$:
\begin{equation}
R(t,s) = \left. \frac{\delta \left\langle \phi(t)\right\rangle}{\delta
j(s)}\right|_{j=0}
 = s^{-1-a} f_R\left(\frac{t}{s}\right)\label{eq:ar}
\end{equation}
where $s$ denotes the start and $t>s$ the observation time,
$j$ is the external conjugate to $\phi$.
This law contains the so-called aging exponent $a$, the universal
scaling function, with the asymptotic behavior
$f_{R}(t/s) \sim (t/s)^{-\lambda_{R}/z}$,
and the autoresponse exponent $\lambda_{R}$.

\gls{lsi} has been shown to describe aging properties of diffusive,
solvable models with $z=2$, like
Acetri~\cite{HenkelDurang2015,Henkel2015_Acetri}, \gls{ew}~\cite{EW}
interface growth and of mean-field like models, exhibiting
long-range interactions~\cite{Henkel2017,HP}.
It also provided agreement with the numerics in case of reaction-diffusion
models \cite{DPLSI,odor2006_isingLSI,henkel2006Aginig}.
However, tests in the critical (1+1)-dimensional contact process
showed systematic deviations in the $t/s \to 1$ limit~\cite{HinDP,Henkel2013}. On
a phenomenological level, these discrepancies could be resolved by the more recent
extension to LLSI~\cite{Henkel2013}, which we shall recall below.

Numerical testing is easier in systems, which do not need to be tuned
to criticality, but exhibit generic scale invariance, like interface models.
For nonequilibrium surface growth dynamics the LLSI predictions
have been found to
be in agreement with the simulations of the $1+1$ dimensional
\gls{kpz} model \cite{PhysRevE.85.030102}. The purpose of the
present study is to extend such investigation to $2+1$ dimensions
in the presence of high precision simulation data available by
simulations of dimer models describing KPZ surface growth \cite{odor09,PhysRevE.84.061150,PhysRevE.89.032146}.

The \gls{kpz} equation \cite{PhysRevLett.56.889} describes the evolution of the height function
$h(\mathbf{x},t)$ in the $d$ dimensional space relative to its mean position
\begin{equation}  \label{KPZ-e}
\partial_t h(\mathbf{x},t) = \nu\nabla^2 h(\mathbf{x},t) +
\lambda(\nabla h(\mathbf{x},t))^2 + \eta(\mathbf{x},t) \ ,
\end{equation}
where $\lambda$ is the amplitude of the up-down anisotropy,
$\nu$ is a smoothing surface tension coefficient and $\eta$
roughening the surface by a zero-mean-value Gaussian noise field exhibiting
the variance
$\langle\eta(\mathbf{x},t)\eta(\mathbf{x^{\prime}},t^{\prime})\rangle =
2 T \nu \delta^d (\mathbf{x-x^{\prime}})(t-t^{\prime})$.
The letter $T$ is related to the noise amplitude (the temperature in
the equilibrium system).

This equation was inspired in part by the stochastic Burgers equation
\cite{burgers74} and can describe the dynamics of simple growth processes
in the thermodynamic limit \cite{H90}, randomly stirred fluids~\cite{forster77},
directed polymers in random media
\cite{kardar85}, dissipative transport \cite{beijeren85,janssen86},
and the magnetic flux lines in superconductors \cite{hwa92}.

The morphology of the surface is usually characterized by the roughness
\begin{align}
 W(L,t) &= \sqrt{\skal{h^2(\mathbf x, t)}_\mathbf x - \skal{h(\mathbf x, t)}_\mathbf x^2}\quad,
 \label{eq:roughness}
 \intertext{where $\skal{\;}_\mathbf x$ denotes an average over all spatial
coordinates.
Simple growth processes are expected to be scale invariant and follow
the Family-Vicsek scaling law~\cite{familyVicsek1985}:}
 W(L,t) &\sim L^{\alpha} f(t / L^z)\quad, \label{eq:familyViczek}
\intertext{with the universal scaling function $f(u)$,}
f(u) &\sim \begin{cases}
 u^\beta & \text{for } u\ll 1 \\
\mathrm{const.} & \text{for } u\gg 1 \ .
\end{cases}\label{eq:kpzScaling}
\end{align}
Here, $\alpha$ is the roughness exponent, describing the stationary state,
where the correlation length exceeds the lateral system size $L$.
The growth regime is governed by the growth exponent $\beta$.
The ratio of these gives the dynamical exponent \mbox{$z=\alpha/\beta$}.
KPZ is invariant to the Galilean symmetry~\cite{forster77},
resulting in the exponent relation
\begin{equation}
 z = 2 / \kla {1 + \beta} \label{eq:galScale}\quad.
\end{equation}

Discrete models set up for KPZ have been studied a lot in the
past decades \cite{meakin,barabasi,krug1997review}. A mapping between
KPZ surface growth in two dimensions and driven lattice gases
has been advanced in~\cite{odor09}.
This is based on the so-called octahedron model, characterized by binary slope
variables at the middle points of the up/down edges. Up slopes in the x or y
directions are represented by $\sigma_{x/y}=1$, while down ones are encoded by
$\sigma_{x/y}=0$.
Thus deposition or removal of octahedra corresponds to a
stochastic cellular automaton with the simple update rules
\begin{equation}\label{rule}
\left(
\begin{array}{cc}
   0 & 1 \\
   0 & 1
\end{array}
\right)
 \overset{p}{\underset{q}{\rightleftharpoons }}
\left(
\begin{array}{cc}
   1 & 0 \\
   1 & 0
\end{array}
\right)
\end{equation}
with probability $p$ ($q$) for attachment (detachment).
By considering edge values to be lattice occupancy variables
we can map the octahedron model onto self-reconstructing
dimers following an oriented migration in the bisection of
$x$ and $y$ directions (see Fig. in \cite{odor09}).
The surface height can be reconstructed for each lattice site from the slope
variables by tracing a path from a reference point: $(1,1)$ with $h_{1,1}:= 0$,
leading to the expression
\begin{equation}\label{height}
h_{i,j} = \sum_{l=1}^i [2\sigma_x(l,1)-1] + \sum_{k=1}^j [2\sigma_y(i,k)-1] \ .
\end{equation}

We have confirmed that this mapping using the parameterization:
$\lambda = 2 p/(p+q)-1$ reproduces the one-point functions of the
continuum model \cite{odor09}. The case $p\approx q$, leading to $\lam\approx0$,
the \gls{ew} model is recovered.
Numerical results for the autocorrelation
have also been found to be in agreement with those of
other KPZ models \cite{PhysRevE.89.032146,halpinhealy2014,carrascoOliviera2014}.
The dimer lattice gas can be studied by very efficient
bit coded simulation methods using graphic cards (GPU) as
detailed in \cite{KONSH2012,PhysRevE.89.032146}.

We performed extensive simulations of the dimer model on lattices with
lateral size of $L=2^{16}$ and periodic boundary conditions. The large systems
serve to stay clear of finite size effects. 
The initial state corresponded to the flat surface and rule~\eqref{rule}
was applied either by a \gls{rs}~\cite{PhysRevE.89.032146} or a sub-lattice
parallel \gls{sca} site selection algorithm.

We calculated the autoresponse function
as described in \cite{PhysRevE.89.032146}.
To introduce a perturbation, we used space-dependent attachment and
detachment probabilities
\begin{equation}
 p_i =
 \begin{cases}
  p_0 + a_i \eps/2 & \text{if }p_0 +a_i\eps/2 \in [0,1] \\
  1 - \eps/2 + a_i \eps/2 & \text{otherwise}
 \end{cases}
\end{equation}
and $q_i = p_0 + q_0 - p_i$,
respectively. Here, $a_i=\pm 1$ and $\eps=0.005$ is a small parameter.
After the waiting time $s$ we used the same stochastic noise $\eta$ (random sequences),
in two realizations. System A evolved, up to the waiting time $s$, with the
site-dependent probabilities $p_i$ and $q_i$ and afterwards with the
uniform ones $p_0$ and $q_0 = 0$.
System B evolved always with spatially uniform attachment and detachment.

From these simulations, we determined the time-integrated response function
\begin{eqnarray}
\lefteqn{ \chi(t,s) = \int_0^s \!\!\D u\: R(t,u) }  \label{Req} \\
&=&
\frac{1}{L^2} \sum_{\vec{r}}^{L^2} \left\langle \frac{h_{\vec{r}}^{(A)}(t,s) -
h_{\vec{r}}^{(B)}(t)}{\eps \Delta}\right\rangle
= s^{-a} f_{\chi}\left( \frac{t}{s} \right)
\quad,\nonumber
\end{eqnarray}
where $a$ is the aging exponent for the autoresponse.
Measurements were performed at exponentially increasing times
\[
t_{i+1} = (t_i  + 10) \cdot \mathrm{e}^m, \quad \text{with} \quad m > 0, \quad t_0 = 0 \quad ,
\]
up to $t_\mathrm{max} = 200 \cdot s$. Throughout this paper
time is measured in \gls{mcs}, defined as one sweep over all lattice sites.

The random-sequential \gls{gpu} implementation from~\cite{PhysRevE.89.032146}
has been modified using a novel combination of the dead border and double tiling
domain-decomposition schemes, which we call DTrDB, in order to
eliminate previously observed correlations.
Details of this algorithm will be discussed elsewhere~\cite{kpzRSUnpub}.
To speed up simulations further, we introduced a \gls{sca} algorithm on \glspl{gpu},
which uses a checkerboard pattern for updates: A \gls{mcs} is performed
by updating all odd sites simultaneously with $p < 1$ and all even sites afterwards
\cite{kellingOdorGemming2016_INES}.
The \gls{gpu} implementations were tested by comparing different schemes.
Direct comparison of the \gls{gpu} results with sequential \gls{cpu} simulations
was impossible on the same level of accuracy, but consistency with former
simulations \cite{PhysRevE.89.032146} could be achieved.

\newcommand\kpzArScaRsRatio{\num{2.08}}

\begin{figure}
 \centering
 \includegraphics{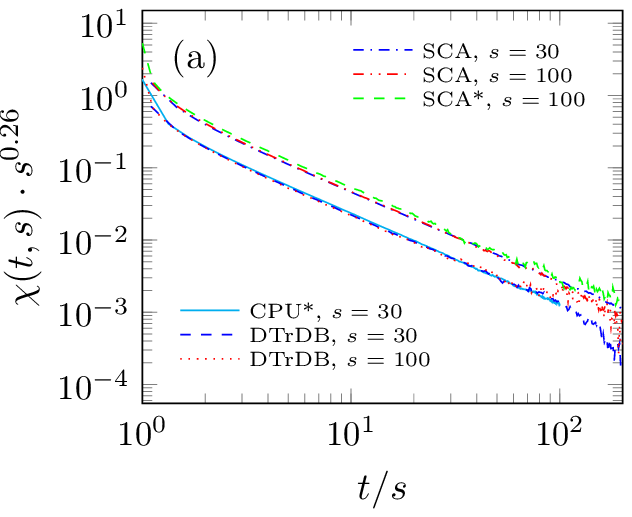}%
 ~
 \includegraphics{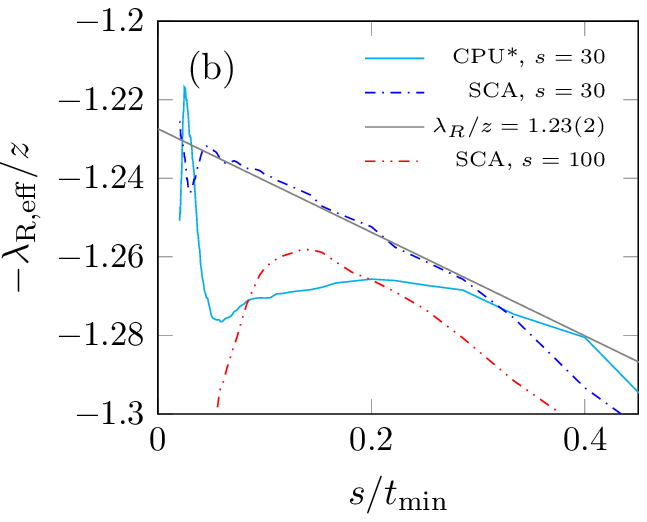}%
 \caption{\label{fig:ar}%
  Simulation results of the integrated height autoresponse,
  comparing variants obtained by \gls{rs} and \gls{sca} simulations.
  (a): Aging collapse of the functions.
  (b): Corresponding effective exponents, extrapolating to
  asymptotic values. Slopes of DTrDB and \gls{sca}* are not
  shown, because the late-time regime was too noisy, due to small sample sizes.
  The black straight line corresponds to a linear fit
  to the \gls{sca} $s=30$ results.
  System and sample sizes are: $L_{\mathrm{\gls{cpu}*}}=2^{13}$,
  $n_{\mathrm{\gls{cpu}*}}=39083$~\cite{PhysRevE.89.032146},
  all others use $L=2^{16}$, with
  $n_{\mathrm{\gls{sca}},s=30}=\NkpzArScaSXXX$,
  $n_{\mathrm{\gls{sca}},s=100}=\NkpzArScaSC$,
  $n_{\mathrm{\gls{sca}*},s=100}=1390$,
  $n_{\mathrm{DTrDB},s=30}=830$ and
  $n_{\mathrm{DTrDB},s=100}=700$.
 }
\end{figure}

Results from various autoresponse calculations are summarized in
Fig.~\ref{fig:ar}(a). The forms of the autoresponse function agree very well
across all types of simulations. The most notable difference is a constant
factor ($\sim\kpzArScaRsRatio$) in the response functions between the \gls{sca}
and \gls{rs} results, which is caused by model-dependent time-scales. Also note
the small shift between \glsname{sca} ($p=\num{.95},q=0$) and \glsname{sca}*
($p=\num{.95},q=\num{.05}$) for $s=100$, caused by the different update
probabilities.

The aging exponent is often determined by performing a manual
collapse of the available datasets for different waiting times $s$. For \gls{rs}
simulations, the value $a_\text{\glsname{rs}}^\text{coll.} = \num{.30}(1)$ was determined
in this way and published in~\cite{PhysRevE.89.032146}. For the \gls{sca}
simulations presented in Fig.~\ref{fig:ar}(a), the value
$a_\text{\glsname{sca}}^\text{coll.}=\num{.26}(1)$ shows the best collapse. However, this
method requires visual inspection of plots to determine for which value of
$a^\text{coll.}$ the data collapse works best, which is prone to bias and
underestimation of the attached error margins.

Numerical computation of the aging exponent involves point-wise division of
autocorrelation functions for different waiting times:
\[
 \frac{\chi(t,s_1)}{\chi(t,s_2)}=
 \frac{s_1^af_{\chi}(t/s_1)}{s_2^af_{\chi}(t/s_2)}
 \overset{(t/s_1=t/s_2)}=
 \kla{\frac{s_1}{s_2}}^a
\]
Since the values $\skal{\chi(t,s)}$ are available only at discrete times an
interpolation is required to compute these ratios at arbitrary $t/s$.  The
simplest option is a linear one, which can also be performed on a
double-logarithmic scale, reducing systematic errors when the interpolation
values follow a power law. In the implicit average over $t$ all points are weighted
with their statistical signal-to-noise ratio, which overall increases the weight
of early times, while in the visual method one is tempted to focus on late times.
The present method yields $a_\text{\gls{sca}}=\num{.24}(2)$, for the \gls{sca}
simulations with $q=0$, and $a_\text{DTrDB}=\num{.27}(2)$, for our new
\gls{rs} simulations with $p=1, q=0$. For comparison, we calculated
$a_\text{\gls{rs}} =\num{.25}(4)$ from the data published
in~\cite{PhysRevE.89.032146}, based on \gls{rs} \gls{cpu} and \gls{gpu}
simulations.
The present data suggest no significant difference between the aging
exponents of \gls{rs} and \gls{sca}.

In order to determine the asymptotic scaling and corrections we
determined (tail) effective exponents
$\lambda_{R,\mathrm{eff}}(t_\mathrm{min})/z$, where each value is the exponent
of a power-law fit to $\chi(t,s)$ in the interval $(t_\mathrm{min},t_\mathrm{max})$
using the form
\begin{equation}
 g_{t_\mathrm{min}}(t) = c \cdot
 t^{-\lambda_{R,\mathrm{eff}}(t_\mathrm{min})/z}\ ,
\end{equation}
with free parameters $c$ and $\lambda_{R,\mathrm{eff}}(t_\mathrm{min})/z$.
The results are displayed in Fig.~\ref{fig:ar}(b) for the three largest
datasets, where $t_\mathrm{min} \lesssim t_\mathrm{max}/4$.
This method suppresses short-wavelength noise but preserves
scaling corrections of larger scales. Only our best dataset
(\gls{sca}, $s=30$) allows a reliable extrapolation for
${\lambda_{R,\mathrm{eff}}(t_\mathrm{min}\to \infty,)/z}$.
The effective exponent curve of the $s=100$ data breaks down
at the end; still the trend observed at early times is in agreement with the
extrapolations for $s=30$. We attribute this to larger oscillations,
similarly as in the case of \gls{cpu} \gls{rs} updates, where, however,
the asymptotic value still appears to agree.

Table~\ref{tab:kpzArEE} summarizes the estimates for the autoresponse
exponent~$\lam_{R}$.
Here we assume $z=1.611(2)$, that can be obtained by the
scaling relation (\ref{eq:galScale}) and using our former,
high precision value $\beta=0.2415(15)$~\cite{PhysRevE.84.061150}.
There is agreement between the results for the considered waiting times
across \gls{rs} and \gls{sca} dynamics.

Considering, that earlier results for the autocorrelation exponent marginally
allow $\lam_C = 2$~\cite{PhysRevE.89.032146}, which is predicted by Krug's
conjecture $\lam_C = d$~\cite{kallabisKrug1999,PhysRevE.55.668}, it is
interesting to note, that \mbox{$\lam_{R}=2$} seems to be satisfied within error
margin. The possible equality $\lam_C =\lam_R$ might point to the existence of a
non-equilibrium fluctuation-dissipation relation in (2+1)-dimensional \gls{kpz}.
The autocorrelation function is defined as
\[
 C(t,s) =
 \skal{\phi(t)\phi(s)}-\skal{
 \phi(t)}\skal{\phi(s)}
 \sim s^{-b} (t/s)^{-\lam_{C}/z}\quad,
\]
with the aging exponent $b=-2\beta$ and definitions analogous to those in
Eq.~\eqref{eq:ar}.
However, one must expect a different relation than in the (1+1)-dimensional
case, because the implied relation for the aging exponents $1+a =
b+2/z$~\cite{PhysRevE.85.030102} does not hold.

\begin{table}
 \centering
 \caption[Height autoresponse exponents]
 {\label{tab:kpzArEE}
  Estimates for the height autoresponse exponent $\lam_{R}$,
  assuming $z=1.611(2)$. Sample and system sizes are listed below
  Fig.~\ref{fig:ar}. Error-margins were estimated visually.
 }
 \centering
 \renewcommand{\arraystretch}{1.2}
 \begin{tabular}{lcc}
  \hline
   &\glsname{cpu}*~\cite{PhysRevE.89.032146} & \glsname{sca} 
  \\
  &$p=\num{.98},q=\num{.02}$ &
  {$p=\num{.95},q=0$}
  \\
  \hline
  $\lam_{R}/z$ &\num{1.25}(3)& \num{1.23}(2) \\
  $\lam_{R}$ &  \num{2.01}(5)& \num{1.98}(4) \\
  \hline
 \end{tabular}
\end{table}


The quality of the available data allows a precise calculation of
effective exponents. Yet, the estimates for the asymptotic values
carry larger uncertainties, due to the unknown corrections to scaling.
Thus, a next step in the \gls{kpz} aging studies is an attempt
to determine these corrections, assuming scaling forms for $\chi_{R}$.
These forms are based on the \gls{lsi} hypothesis.
For the time-integrated autoresponse, Eq.~\eqref{Req},
\gls{lsi} theory for KPZ predicts the scaling function
\begin{align}
 f_{\chi,\mathrm{LSI}}(t/s) &= A_0 (t/s)^{-\lam_R/z}\kla{1-s/t}^{-1-a'}\quad,
 \label{eq:ArLsi}
\end{align}
where $A_0$ is a normalization factor and $a'$ is expected to be another universal
exponent, like the aging exponent $a$. A different form, adding logarithmic
corrections was proposed recently in~\cite{Henkel2013}:
\begin{align}
 f_{\chi,\mathrm{L^2LSI}} &=
 (t/s)^{1-\lam_R/z}\left[A_0\kla{1-(1-s/t)^{-a'}}\right.\nonumber\\
 \label{eq:ArL2Lsi}
&+\left. (1-s/t)^{-a'}\cdot\kla{A_1\ln(1-s/t) + A_2\ln^2(1-s/t)}\right]\quad,
\end{align}
where the sum of logarithmic terms to second order results from the
assumption, that the primary field $\phi$ of the system is replaced by a
doublet and the scaling dimensions are represented by $2\times2$ matrices. The
solution in~\cite{Henkel2013} uses a Schrödinger-invariant Lie algebra, which
implies $z=2$, a dynamical exponent different from that of the \gls{kpz}
universality class. However, the scaling form \eqref{eq:ar} is invariant of the
value of $z$, because it depends on the ratio $\lambda_R/z$ only. It should be
noted, that, this does not hold for the space-dependent part of the response
function, which remains an open problem.

The scaling function~\eqref{eq:ArL2Lsi} resembles a form, which contains
the first two lowest order correction terms of a logarithmic series to~\eqref{eq:ArLsi}.
We shall test by fitting if, an assumed more generalized power series form
\begin{align}
 f_{\chi,\mathrm{L^JLSI}} &=
 (t/s)^{1-\lam_R/z}\left[A_0\kla{1-(1-s/t)^{-a'}}\right.\nonumber\\
 \label{eq:ArLLsi}
 &+\left. (1-s/t)^{-a'}\cdot\sum\limits_{j>0}^J A_j\ln^j(1-s/t)\right] \quad,
\end{align}
  which, given enough terms, might fit a broad range of data, really supports
  the expected L\textsuperscript2LSI theory with the scaling~\eqref{eq:ArL2Lsi}.
However, an LSI extension with triplets, or beyond, would also give physical
meaning to some terms with $j\geq3$. Thus these terms being relevant to describe
the data would point to the necessity of higher orders in the extension of LSI.

\begin{figure}
 \centering
 \includegraphics{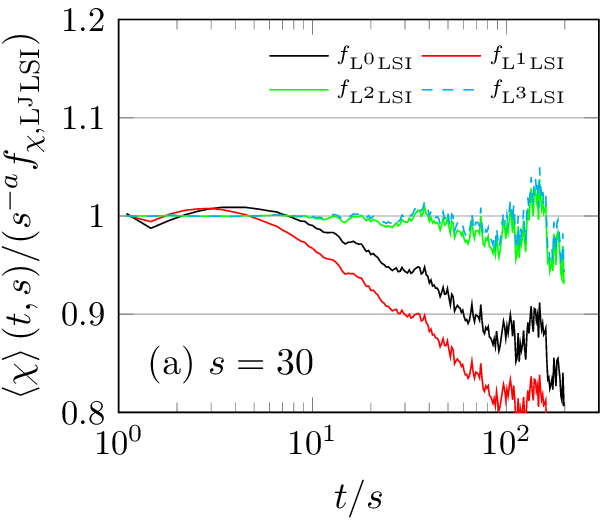}%
 ~
 \includegraphics{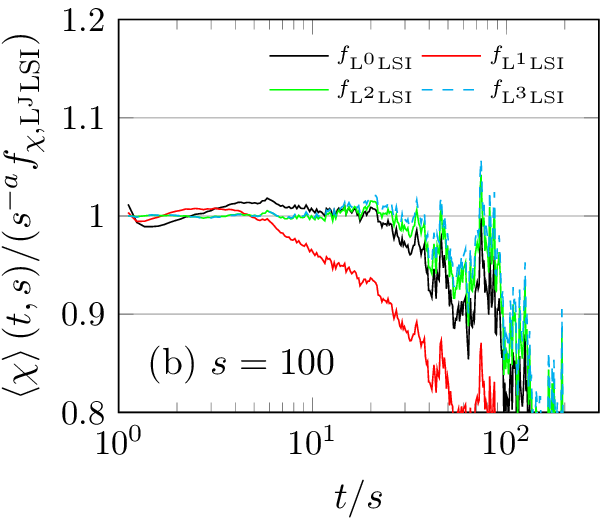}%
 \caption
 {\label{fig:llsi}
  Plots of Eq.~\eqref{eq:ArLsiRatio} in case of \gls{sca} autoresponse
  calculations with $p=\num{.95}$ and $q=0$. Sample sizes are
  $n_{\mathrm{\gls{sca}},s=30}=\NkpzArScaSXXX$ for $s=30$ (a) and
  $n_{\mathrm{\gls{sca}},s=100}=\NkpzArScaSC$ for $s=100$ (b).
  Best fits are determined from the region $1\leq t/s\leq10$.
 }
\end{figure}

We have tested different ($J$) levels of the series (\ref{eq:ArLLsi}) with
our data, obtained from the most precise \gls{sca} simulations.
Figure~\ref{fig:llsi} shows plots of the ratio of data and best fit.
This is a visual representation of how well forms for $J\in[0,3]$ describe the data:
\begin{equation}
 \frac{\skal{\chi}(t/s)}{s^{-a}f_{\chi,\mathrm{L^JLSI}}(t/s)}
 \imeq 1 \qquad\text{for }t/s>1 \quad.
 \label{eq:ArLsiRatio}
\end{equation}

\begin{table}[bt!]
 \caption{\label{tab:llsi}
  Parameters for best fits of $f_{\chi,\mathrm{L^JLSI}}$ forms to \gls{kpz}
  autoresponse functions for $1\leq t/s\leq200$. Values for
  $\lam_{R}/z$ in parenthesis result from fits considering $q\leq t/s
  \leq 10$, as presented in figure~\ref{fig:llsi}. $a=\num{.24}$ for all fits.
  Error margins are
  not given, because the method employed for fitting does not provide meaningful
  estimates.
 }
 \centering
 \begin{tabular}{llrrrrrr}
  \hline
  && $\lam_{R}/z$ & $a'$ & $A_0$ & $A_1$ & $A_2$ &$A_3$ \\\hline
  \multirow{4}{*}{\rotatebox{90}{$s=30$}}
  &$f_\mathrm{L^0LSI}$ & \num{1.164} (\num{1.167})& \num{0.016} & \num{38.833} \\
  &$f_\mathrm{L^1LSI}$ & \num{1.164} (\num1.144{}) & \num{0.023} & \num{35.085}
  & \num{0.187} \\
  &$f_\mathrm{L^2LSI}$ & \num{1.224} (\num1.219{}) & \num{0.501}
  & \num{4.938} & \num{1.772} & \num{-0.431} \\
  &$f_\mathrm{L^3LSI}$ & \num{1.224} (\num1.224{}) & \num{0.505} & \num{4.790} &
  \num{1.716} & \num{-0.422} & \num{-0.004} \\
  \hline
  \multirow{4}{*}{\rotatebox{90}{$s=100$}}
  &$f_\mathrm{L^0LSI}$ & \num{1.186} (\num{1.191}) & \num{0.006} & \num{102.584} \\
  &$f_\mathrm{L^1LSI}$ & \num{1.165} (\num{1.142}) & \num{0.100} & \num{14.444}
  & \num{0.844} \\
  &$f_\mathrm{L^2LSI}$ & \num{1.230} (\num{1.224}) & \num{0.490} & \num{5.544}
  & \num{2.019} & \num{-0.472} \\
  &$f_\mathrm{L^3LSI}$ & \num{1.230} (\num{1.233}) & \num{0.475} & \num{5.506} &
  \num{1.914} & \num{-0.437} & \num{-0.008} \\
  \hline
 \end{tabular}
\end{table}

Non-linear fits for $J>0$ do not converge using the classical least-squares
Levenberg--Marquardt algorithm~\cite{Levenberg1944,Marquardt1963}. To obtain
the parameters presented in table~\ref{tab:llsi}, the Nelder-Mead
method~\cite{NelderMead1965} was employed, which does not provide statistical
error estimates for the fit parameters.  Fit results can be governed by any
of a multitude of local
minima, depending on the initial guesses and the choosen fit interval. Judging
by the connected variation in parameter values, the accuracy of the tabulated
parameters should be assumed to be no better than~\SI{20}\%, except for the
values of $\lam_{R}/z$ which vary by less than~\SI5\%.

It is apparent from figure~\ref{fig:llsi} that the uncorrected \gls{lsi} ansatz
fails to describe the asymptotic behavior of $\chi$, giving
$\lam_{R}/z\approx \num{1.17}$. So does the logarithmic form with $J=1$.
The form with $J=2$, which is predicted by the theory yields much better fits,
with $\lam_{R}/z\approx \num{1.22}$, agreeing with the asymptotic
value obtained earlier
$\lam_{R}^\mathrm{tail}/z = \num{1.23}(2)$.
The parameter fits presented in table~\ref{tab:llsi} take into account the
observed time interval $1\leq t/s\leq200$. When the fit is limited to the interval $1\leq
t/s\leq10$, the results for $\lam_{R}$ (values in parentheses) do not
change significantly. This means, that the $f_{\chi,\mathrm{L^2LSI}}$ form describes the
corrections, affecting the autoresponse function at early times, well enough to
determine the correct asymptotic autoresponse exponent just using early-time
data.

The form with $J=3$ shows marginally better agreement with the data in
figure~\ref{fig:llsi}. In fits to the whole observed time interval, the
amplitude $A_3$ of the added third-order term is severely suppressed
(table~\ref{tab:llsi}). Adding another fit parameter, a slightly better fit
would be expected. The small absolute value of $A_3$ in relation to $A_2$ suggests,
that a third order correction does not carry physical meaning, supporting the
L\textsuperscript{2}\gls{lsi} theory.

The values of the coefficients for $J=2$ and $3$ are similar at different
waiting times. This satisfies our expectation, since aging is described by the
$s^{-a}$ term in equation~\eqref{Req} alone and the functional form of
$f_\chi(t/s)$ should not depend on $s$ explicitly.
The autoresponse functions we obtained by less precise simulations also
agree with the L\textsuperscript2LSI theory, but they exhibit too much
noise to exclude a logarithmic series like~\eqref{eq:ArLLsi}.

In conclusion, we provide numerical evidence that the L\textsuperscript2LSI theory
describes well aging data of the autoresponse function for all measured
times in case of the $2+1$ dimensional KPZ surface growth.
We obtained precise estimates for the autoresponse exponent as well as
for the aging exponents. In particular a $\lambda_R = 2.00(6)$ estimate
seems to emerge from our high precision parallel simulations.
Our code can be extended to also calculate the space-dependent part of the
\gls{kpz} response function.
For the autocorrelation functions of the \gls{kpz} model
$f_{C,\mathrm{L^2LSI}}$ a form is yet to be proposed. Our \gls{gpu}
simulations generate high precision correlation data for heights
as well as density variables that remains to be tested later
against different aging functions~\cite{kpzRSUnpub}.

\vspace*{1em}

\noindent
{\bf Acknowledgments:}\\
We thank M.~Henkel for helpful discussions and comments.
Support from the Hungarian research fund OTKA (Grant No.~K109577), the
Initiative and Networking Fund of the Helmholtz Association via the W2/W3
Program \mbox{(W2/W3-026)} and the International Helmholtz Research School
NanoNet \mbox{(VH-KO-606)} is acknowledged.
We gratefully acknowledge computational
resources provided by the HZDR computing center, NIIF Hungary and the Center for
Information Services and High Performance Computing (ZIH) at TU Dresden.

 \bibliography{./bib}
\bibliographystyle{iopart-num}

\end{document}